\documentclass[prd,reprint,showpacs,showkeys]{revtex4-1}
\usepackage{amssymb}
\usepackage{amsmath}
\usepackage{graphicx}
\usepackage[font={footnotesize,it}]{caption}

\pdfoutput=1

\begin{document}

\title[ ]{Quantum probes of timelike naked singularities in $2+1-$%
dimensional power - law spacetimes}
\author{O. Gurtug}
\email{ozay.gurtug@emu.edu.tr}
\author{M. Halilsoy}
\email{mustafa.halilsoy@emu.edu.tr}
\author{S. Habib Mazharimousavi}
\email{habib.mazhari@emu.edu.tr}

\begin{abstract}
The formation of naked singularities in $2+1-$ dimensional power - law
spacetimes in linear Einstein-Maxwell and Einstein-scalar theories sourced
by azimuthally symmetric electric field and a self-interacting real scalar
field respectively, are considered in view of quantum mechanics. Quantum
test fields obeying the Klein-Gordon and Dirac equations are used to probe
the classical timelike naked singularities developed at $r=0$. We show that
when the classically singular spacetimes probed with scalar waves, the
considered spacetimes remains singular. However, the spinorial wave probe of
the singularity in the metric of a self-interacting real scalar field
remains quantum regular. The notable outcome in this study is that the
quantum regularity/singularity can not be associated with the energy
conditions.
\end{abstract}

\maketitle
\affiliation{Department of Physics, Eastern Mediterranean University, Gazima\u{g}usa,
Turkey. }

\section{Introduction}

In recent years, general relativity in $2+1-$ dimensions has been one of the
attractive arena for understanding the general aspects of black holes
physics. The main motivation of this attraction is the existing tractable
mathematical structure when compared with the higher dimensional
counterparts. The preliminary works in this field was popularized by the
black hole solution of Banados-Teitelboim-Zanelli (BTZ), a spacetime sourced
by a negative cosmological constant \cite{1,2,3}. Extension of $2+1-$
dimensional solutions to Einstein-Maxwell (EM) cases followed in \cite{4}
and its Massive Gravity version is given in \cite{5}. The static and
rotating charged black hole in $2+1-$dimensional Brans-Dicke theory was
studied in \cite{6} and rotating black holes with torsion was considered in 
\cite{7}. The $2+1-$dimensional charged black hole with nonlinear
electrodynamic coupled to gravity has been studied in \cite{8} and with a
scalar hair has been given in \cite{9}. The peculiar feature in the
aforementioned studies in the EM theory is that the electric field is
considered in the radial direction.

In recent decades, the physical properties of the solutions presented in
both linear and nonlinear electromagnetism have been investigated by
researchers. The solutions admitting black holes are analyzed in terms of
thermodynamical aspects such as temperature and entropy \cite{3, T2, T3}.
Furthermore, the AdS/CFT correspondence which relates thermal properties of
black holes in the AdS space to a dual CFT is another important achievement
of $2+1-$dimensional gravity \cite{10}.

On the other hand, the solutions admitting naked singularities are not
analyzed in details and hence, it requires further care as far as the cosmic
censorship hypothesis is concerned. Therefore, the resolution of
singularities becomes important not in $3+1-$dimensional gravity, but also
in both lower and higher dimensional gravity as well. Because of the scales
where these singularities are forming, their resolution requires a
consistent theory at these small scales. The theory of quantum gravity seems
to be the most promising theory, however, it is still "under construction".
An alternative method for resolving the singularities is proposed by
Horowitz and Marolf (HM) \cite{11} by developing the work of Wald \cite{12}.
According to this method, the classical notion of a curvature singularity
that is regarded as \textit{geodesics incompleteness} with respect to point
particle probe is replaced by \textit{quantum singularity} with respect to
wave probes.

In this paper, our focus will be on the $2+1-$dimensional power-law
spacetimes. Spherically symmetric power-law metrics for dimensions $n\geq 4$
has been investigated in view of quantum mechanics by Blau, Frand and Weiss
(BFW) in \cite{Y1}, by employing the method of HM. The formation of naked
singularities in 2-parameter family of 
\begin{equation}
ds^{2}=\eta x^{p}\left( -dx^{2}+dy^{2}\right) +x^{q}d\Omega _{d}^{2},
\end{equation}%
Szekeres - Iyer \cite{Y2,Y3,Y4}, metrics is probed with scalar field. It has
been shown that the timelike naked singularity at $x=0$ for $\eta =-1$
satisfying the Dominant Energy Condition (DEC) are quantum mechanically
singular in the sense of the HM criterion.

Another study in line with power - law spacetimes was considered by Helliwel
and Konkowski (HK) in \cite{Y5}. HK considered cylindrically symmetric four
- parameter power - law metrics in $3+1-$dimension in the form of%
\begin{equation}
ds^{2}=-r^{\alpha }dt^{2}+r^{\beta }dr^{2}+\frac{1}{C^{2}}r^{\gamma }d\theta
^{2}+r^{\delta }dz^{2},
\end{equation}%
in the limit of small $r,$ where $\alpha ,\beta ,\gamma ,\delta $ and $C$
are constant parameters. HK classified the metric (2) as Type I, if $\alpha
=\beta ,$ and given by%
\begin{equation}
ds^{2}=r^{\beta }\left( -dt^{2}+dr^{2}\right) +\frac{1}{C^{2}}r^{\gamma
}d\theta ^{2}+r^{\delta }dz^{2}.
\end{equation}%
According to the analysis of HK, a large set of classically singular
spacetimes emerges quantum mechanically non-singular, if it is probed with
scalar waves having non - zero azimuthal quantum number $m$ and axial
quantum number $k$ in the sense of HM criterion. HK have also argued the
possible relation with the energy conditions that can be used to eliminate
the quantum singular spacetimes.

In $2+1-$dimensional gravity, the method of HM has been used in the
following works to probe the timelike naked curvature singularities: The BTZ
black hole is considered in \cite{13}. The EM extension of BTZ black hole
both in linear and nonlinear theory and in Einstein- Maxwell -dilaton theory
is considered in \cite{14}. The formation of naked singularities for a
magnetically charged solution in Einstein-power-Maxwell-theory is considered
in \cite{15}. Occurrence of naked singularities in Einstein-nonlinear
electrodynamics with circularly symmetric electric field is considered in 
\cite{16}. In these studies, the timelike naked singularity is probed with
waves that differs in spin structure. Namely, the bosonic and the fermionic
waves are used that obey the Klein-Gordon and the Dirac equation,
respectively. The common outcome in these studies is that the naked
singularity remains quantum singular when it is probed with bosonic waves.
However, probing the singularity with fermionic waves has revealed that only
the magnetically charged solution in Einstein-power-Maxwell theory is
singular. The other spacetimes considered so far behaves quantum regular
against fermionic waves.

To our knowledge, the analysis of power - law metrics in $2+1-$dimensions
has not been considered so far. This fact will be the main motivation for
the present study. The solutions admitting timelike naked singularities in
the linear Einstein - Maxwell (EM) \cite{17} and Einstein - scalar (ES) \cite%
{Y6} theories sourced by azimuthally symmetric electric field and a
self-interactin real scalar field, respectively, will be investigated within
the framework of quantum mechanics. The peculiar feature of both solutions
is that they admit metrics in power - law form in $2+1-$dimensional gravity.

The solution in linear EM theory with azimuthally symmetric electric field
in $2+1-$dimension was given in \cite{17}. To our knowledge, the singularity
structure of the solution presented in \cite{17} has not been studied so
far. We are aiming in this study to investigate the solution admitting naked
singularity in \cite{17}. In our analysis, the classical naked singularity
will be probed with quantum fields obeying the massless Klein-Gordon and
Dirac equations. We showed that against both probes the spacetime remains
quantum mechanically singular. This happens in spite of the fact that the
weak, strong and dominant energy conditions are manifestly satisfied.

The electric field component in EM extensions was considered to be radial so
far while the possibility of a circular electric field went unnoticed.
Recall from the Maxwell equations, $\mathbf{\nabla \times E}\sim \frac{%
\partial \mathbf{B}}{\partial t}$ and $\mathbf{\nabla \times B\sim }\frac{%
\partial \mathbf{E}}{\partial t}$ that the possibility of a constant $%
\mathbf{E}$ ($=E_{0}=F_{t\theta }=$ constant) and gradient form of $\mathbf{%
B,}$ (i.e. $\mathbf{B}=\mathbf{\nabla }b,$ \ for $b$ a scalar function,
independent of time) may occur. When confined to $2+1-$dimensions such $%
\mathbf{E}$ and $\mathbf{B}$ satisfy Maxwell's equations trivially with
singularity due to a physical source at $r=0$. It is this latter case that
we wish to point out and investigate in this paper.

The paper is organized as follows. In section II, the solutions obtained in 
\cite{17} and in \cite{Y6} are reviewed and the structure of the resulting
spacetimes are briefly introduced. In section III, the definition of quantum
singularity is summarized and the timelike naked singularity in the
considered spacetimes are analyzed with quantum fields obeying the
Klein-Gordon and Dirac equations. The paper is concluded with a conclusion
in section IV.

\section{Review of the solutions admitting power - law metrics in $\left(
2+1\right) -$dimension}

\subsection{Rederivation of the linear Einstein - Maxwell solution with
azimuthally symmetric electric field}

We start with the Einstein-Maxwell action given by 
\begin{equation}
I=\frac{1}{2}\int d^{3}x\sqrt{-g}\left( R-\mathcal{F}\right) .
\end{equation}%
in which $R$ is the Ricci scalar and $\mathcal{F}=F_{\mu \nu }F^{\mu \nu }$
is the Maxwell invariant with $F_{\mu \nu }=\partial _{\mu }A_{\nu
}-\partial _{\nu }A_{\mu }$. The circularly symmetric line element is given
by 
\begin{equation}
ds^{2}=-A\left( r\right) dt^{2}+\frac{1}{B(r)}dr^{2}+r^{2}d\theta ^{2},
\end{equation}%
where $A(r)$ and $B(r)$ are unknown functions of $r$ and $0\leq \theta \leq
2\pi $. The electric field ansatz is chosen to be normal to radial direction
and uniform i.e., 
\begin{equation}
\mathbf{F}=E_{0}dt\wedge d\theta
\end{equation}%
in which $E_{0}=$constant \cite{16}. The dual field is found as $^{\star }%
\mathbf{F}=\frac{E_{0}}{r}\sqrt{\frac{B}{A}}dr.$ It is known that the
integral of $^{\star }\mathbf{F}$ gives the total charge. Let us note that
even in a flat space with $A=B=1$ we obtain a logarithmic expression for the
charge, i.e. $Q\left( r\right) \sim \ln r$. This electric field is derived
from an electric potential one-form given by%
\begin{equation}
\mathbf{A}=E_{0}\left( a_{0}td\theta -b_{0}\theta dt\right) ,
\end{equation}%
in which $a_{0}$ and $b_{0}$ are constants satisfying $a_{0}+b_{0}=1.$ The
Maxwell's equation%
\begin{equation}
d\left( ^{\star }\mathbf{F}\right) =0,
\end{equation}%
is trivially satisfied. Note that the invariant of electromagnetic field is
given by 
\begin{equation}
\mathcal{F}=2F_{t\theta }F^{t\theta }=\frac{-2E_{0}^{2}}{A\left( r\right)
r^{2}}.
\end{equation}%
Next, the Einstein-Maxwell equations are given by 
\begin{equation}
G_{\mu }^{\nu }=T_{\mu }^{\nu }
\end{equation}%
in which 
\begin{equation}
T_{\ \nu }^{\mu }=-\frac{1}{2}\left( \delta _{\ \nu }^{\mu }\mathcal{F}%
-4F_{\nu \lambda }F^{\mu \lambda }\right) .
\end{equation}%
Having $\mathcal{F}$ known one finds%
\begin{equation}
T_{\ t}^{t}=T_{\ \theta }^{\theta }=\frac{1}{2}\mathcal{F},
\end{equation}%
and 
\begin{equation}
T_{\ r}^{r}=-\frac{1}{2}\mathcal{F}
\end{equation}%
as the only non-vanishing energy-momentum component. To proceed further, we
must have the exact form of the Einstein tensor components given by%
\begin{equation}
G_{t}^{t}=\frac{B^{\prime }}{2r}
\end{equation}%
\begin{equation}
G_{r}^{r}=\frac{BA^{\prime }}{2rA}
\end{equation}%
and%
\begin{equation}
G_{\theta }^{\theta }=\frac{2A^{\prime \prime }AB-A^{\prime 2}B+A^{\prime
}B^{\prime }A}{4A^{2}},
\end{equation}%
in which a 'prime' means $\frac{d}{dr}.$ The field equations then read as
follows%
\begin{equation}
\frac{B^{\prime }}{2r}=\frac{1}{2}\mathcal{F},
\end{equation}%
\begin{equation}
\frac{BA^{\prime }}{2rA}=-\frac{1}{2}\mathcal{F}
\end{equation}%
and%
\begin{equation}
\frac{2A^{\prime \prime }AB-A^{\prime 2}B+A^{\prime }B^{\prime }A}{4A^{2}}=%
\frac{1}{2}\mathcal{F}.
\end{equation}%
The above field equations admit the following solutions for $A$ and $B:$%
\begin{equation}
A=\frac{1}{B}=\chi r^{2E_{0}^{2}}
\end{equation}%
in which $\chi >0$ is an integration constant and without loss of generality
we set it to $\chi =1$. Hence the line element becomes%
\begin{equation}
ds^{2}=r^{2E_{0}^{2}}\left( -dt^{2}+dr^{2}\right) +r^{2}d\theta ^{2}.
\end{equation}%
This is a black point solution with the horizon at the origin which is the
singular point of the spacetime with Kretschmann scalar%
\begin{equation}
\mathcal{K}=\frac{12E_{0}^{4}}{r^{4\left( 1+E_{0}^{2}\right) }}.
\end{equation}%
The solution has a single parameter which is the electric field $E_{0}.$
Setting $E_{0}=0$ makes the solution the $\left( 2+1\right) -$dimensional
flat spacetime. Note also that for the choice $E_{0}=1,$ from (21), we
obtain a conformally flat metric with conformal factor $r^{2}.$ It is
observed that the strength of $E_{0}$ serves to increase the degree of
divergence of the scalar curvature. Based on our energy momentum tensor
components one finds that the energy density and the radial and tangential
pressures are given by%
\begin{equation}
\rho =-T_{t}^{t}=\frac{E_{0}^{2}}{r^{2\left( 1+E_{0}^{2}\right) }},\text{ \ }
\end{equation}%
\begin{equation}
\text{\ }p=T_{r}^{r}=\rho =\frac{E_{0}^{2}}{r^{2\left( 1+E_{0}^{2}\right) }},
\end{equation}%
and 
\begin{equation}
q=T_{\theta }^{\theta }=-\rho =-\frac{E_{0}^{2}}{r^{2\left(
1+E_{0}^{2}\right) }}.
\end{equation}%
Therefore the weak energy conditions (WEC) i.e., i) $\rho \geq 0,$ ii) $\rho
+p\geq 0$ and iii) $\rho +q\geq 0$ all are satisfied. The strong energy
conditions are also satisfied i.e., the WECs together with iv) $\rho
+p+q\geq 0.$ Dominant Energy Condition (DEC) i.e. $p_{eff}\geq 0$ and
Causality Condition (CC) i.e. $0\leq p_{eff}\leq 1$ are also easily
satisfied knowing that $p_{eff}=\frac{p+q}{2}=0.$

\subsection{Exact radial solution to $\left( 2+1\right) $ $-$dimensional
gravity coupled to a self-interacting real scalar field}

Exact radial solution with a self-interacting, real, scalar field coupled to
the $\left( 2+1\right) $ $-$dimensional gravity is given by Schmidt and
Singleton in \cite{Y6}. The action representing $\left( 2+1\right) $ $-$%
dimensional gravity with a self-interacting scalar field is given by%
\begin{equation}
I=\frac{1}{2\kappa }\int d^{3}x\sqrt{-g}\left( R+\mathcal{L}_{S}\right)
\end{equation}%
in which $\kappa =8\pi G$ is the coupling constant, $G$ \ denotes the
Newton's constant, $R$ Ricci scalar and $\mathcal{L}_{S}$ represents the
Lagrangian of the self-interacting scalar field $\phi $ given by%
\begin{equation}
\mathcal{L}_{S}=-2\partial _{\mu }\phi \partial ^{\mu }\phi -V\left( \phi
\right) .
\end{equation}%
The potential $V\left( \phi \right) $ is a Liouville potential described by%
\begin{equation}
V\left( \phi \right) =e^{-\sqrt{\kappa }\phi }.
\end{equation}%
With reference to \cite{Y6}, the metric and the scalar field are found as,%
\begin{equation}
ds^{2}=-Kr^{2}dt^{2}+dr^{2}+r^{2}d\theta ^{2},
\end{equation}

\begin{equation}
\phi \left( r\right) =\frac{1}{\sqrt{\kappa }}\ln \left( r\right) ,
\end{equation}%
where $K$ is a constant parameter. The only non vanishing energy - momentum
tensor component is the radial presure componenet, $T_{rr}=\frac{1}{\kappa
r^{2}}.$The Kretschmann scalar is given by $\mathcal{K}=\frac{4}{r^{4}},$%
which indicates a scalar curvature singularity at $r=0$. This solution has
no horizon and hence, the singularity at $r=0$ is a timelike naked
singularity.

According to the energy - momentum tensor components one finds that the
energy density and the radial and tangential pressures are given by;%
\begin{equation}
\rho =0,\text{ \ \ }p=\frac{1}{r^{2}},\text{ \ \ \ }q=0.
\end{equation}%
As a consequence, the weak energy conditions (WEC) i.e., i) $\rho \geq 0,$
ii) $\rho +p\geq 0$ and iii) $\rho +q\geq 0$ , the strong energy conditions
(SEC) i.e., the WECs together with iv) $\rho +p+q\geq 0$ are all satisfied.
The DEC, $p_{eff}=\frac{p+q}{2}\geq 0$ is also satisfied.

\section{Singularity Analysis}

It has been known that the spacetime singularities inevitably arise in the
Einstein's theory of relativity. It describes the "end point" or incomplete
geodesics for timelike or null trajectories followed by classical particles.
Among the others, naked singularities which is visible from outside needs
further care as far as the weak cosmic censorship hypothesis is concerned.
It is believed that, naked singularity forms a threat to this hypothesis. As
a result of this, understanding and the resolution of naked singularities
seems to be extremely important for the deterministic nature of general
relativity. The general belief in the resolution of the singularities is to
employ the methods imposed by the quantum theory of gravity. However, the
lack of a consistent quantum gravity leads the researchers to alternative
theories in this regard. String theory \cite{18,19}and loop quantum gravity 
\cite{20} constitute two major study fields in resolving singularities.
Another alternative method; following the work of Wald \cite{12}, was
proposed by Horowitz and Marolf (HM) \cite{11}, which incorporates the
"self-adjointness" of the spatial part of the wave operator. Hence, the
classical notion of \textit{geodesics incompleteness} with respect to
point-particle probe will be replaced by the notion of \textit{quantum
singularity} with respect to wave probes.

In this paper, the method proposed by HM will be used in analyzing the naked
singularity. This method in fact has been used successfully in $\left(
3+1\right) $ and higher dimensional spacetimes. The complete list in these
spacetimes given in \cite{21,22,23,24,25,26,27,28,29,30,31,32,33,34,35}. The
main purpose in these studies is to understand whether these classically
singular spacetimes turn out to be quantum mechanically regular if they are
probed with quantum fields rather than classical particles.\ The idea is in
analogy with the fate of a classical atom in which the electron should
plunge into the nucleus but rescued with quantum mechanics. The main
concepts of this method which can be applied only to static spacetimes
having timelike singularities is summarized briefly as follows.

Let us consider, the Klein-Gordon equation for a free particle that
satisfies $i\frac{d\psi }{dt}=\sqrt{\mathcal{A}_{E}}\psi ,$ whose solution
is $\psi \left( t\right) =\exp \left[ -it\sqrt{\mathcal{A}_{E}}\right] \psi
\left( 0\right) $ in which $\mathcal{A}_{E}$ denotes the extension of the
spatial part of the wave operator. If the wave operator $\mathcal{A}$ is not
essentially self-adjoint, in other words if $\mathcal{A}$ has an extension,
the future time evolution of the wave function $\psi \left( t\right) $ is
ambiguous. Then, the HM method defines the spacetime as quantum mechanically
singular. However, if there is only a single self-adjoint extension, the
wave operator $\mathcal{A}$ is said to be\ essentially self-adjoint and the
quantum evolution described by $\psi \left( t\right) $ is uniquely
determined by the initial conditions. According to the HM method, this
spacetime is said to be quantum mechanically nonsingular. The essential
self-adjointness of the operator $\mathcal{A}$, can be verified by using the
deficiency indices and the Von Neumann's theorem that considers the
solutions of the equation%
\begin{equation}
\mathcal{A}^{\ast }\psi \pm i\psi =0,
\end{equation}%
and showing that the solutions of (32), do not belong to Hilbert space $%
\mathcal{H}$. (We refer to; \cite{21,36,37,38} for detailed mathematical
analysis.)

\subsection{Quantum probes of the linear Einstein - Maxwell solution with
azimuthally symmetric electric field}

\subsubsection{Klein-Gordon Fields}

The massless Klein-Gordon equation for a scalar wave can be written as

\begin{equation}
\frac{1}{\sqrt{-g}}\partial _{\mu }\left[ \sqrt{-g}g^{\mu \nu }\partial
_{\nu }\right] \psi =0.
\end{equation}%
The Klein-Gordon equation can be written for the metric (21), by splitting
the temporal and spatial part as

\begin{equation}
\frac{\partial ^{2}\psi }{\partial t^{2}}=\frac{\partial ^{2}\psi }{\partial
r^{2}}+\frac{1}{r}\frac{\partial \psi }{\partial r}+r^{2(E_{0}^{2}-1)}\frac{%
\partial ^{2}\psi }{\partial \theta ^{2}}.
\end{equation}%
This equation can also be written as%
\begin{equation}
\frac{\partial ^{2}\psi }{\partial t^{2}}=-\mathcal{A}\psi ,
\end{equation}%
where $\mathcal{A}$ is the spatial operator given by 
\begin{equation}
\mathcal{A}=-\frac{\partial ^{2}}{\partial r^{2}}-\frac{1}{r}\frac{\partial 
}{\partial r}-r^{2(E_{0}^{2}-1)}\frac{\partial ^{2}}{\partial \theta ^{2}},
\end{equation}%
and according to the HM method, it is subjected to be investigated whether
its self-adjoint extensions exists or not. This is achieved by assuming a
separable solution to equation (32) in the form of $\psi (r,\theta
)=R(r)Y(\theta ),$ which yields the radial equation as%
\begin{equation}
\frac{\partial ^{2}R\left( r\right) }{\partial r^{2}}+\frac{1}{r}\frac{%
\partial R\left( r\right) }{\partial r}+\left( cr^{2(E_{0}^{2}-1)}\pm
i\right) R\left( r\right) =0.
\end{equation}%
with $c$ the separation constant. The essential self-adjointness of the
spatial operator $\mathcal{A}$ requires that neither of the two solutions of
the above equation is square integrable over all space $L^{2}(0,\infty ).$%
The square integrability of the solution of the above equation for each sign 
$\pm $ is checked by calculating the squared norm of the obtained solution
in which the function space on each $t=$ constant hypersurface $\Sigma $ is
defined as $\mathcal{H=}\{R|\left\Vert R\right\Vert <\infty \}.$ The squared
norm for the metric (2) is given by,

\begin{equation}
\left\Vert R\right\Vert ^{2}=\int_{0}^{\infty }\frac{\left\vert R_{\pm
}\left( r\right) \right\vert ^{2}r}{\sqrt{A(r)B(r)}}dr.
\end{equation}%
The squared norm is investigated for three different cases of the value of
electric field $E_{0}.$

\paragraph{Case 1:$\ $ $E_{0}^{2}=1:$}

For this particular case, equation (37) transforms to%
\begin{equation}
\frac{\partial ^{2}R\left( r\right) }{\partial r^{2}}+\frac{1}{r}\frac{%
\partial R\left( r\right) }{\partial r}+\left( c\pm i\right) R\left(
r\right) =0,
\end{equation}%
whose solution is given by%
\begin{equation}
R(r)=a_{1}J_{0}(\sqrt{\hat{c}}r)+a_{2}N_{0}\left( \sqrt{\hat{c}}r\right) ,
\end{equation}%
such that, $\hat{c}=c\pm i$.

\paragraph{Case 2: $0<E_{0}^{2}<1:$}

In this case, equation (37) becomes,%
\begin{equation}
\frac{\partial ^{2}R\left( r\right) }{\partial r^{2}}+\frac{1}{r}\frac{%
\partial R\left( r\right) }{\partial r}+\frac{c}{r^{2(1-E_{0}^{2})}}R\left(
r\right) =0,
\end{equation}%
and the solution is given by,%
\begin{equation}
R(r)=a_{3}J_{0}(\frac{\sqrt{c}}{E_{0}^{2}}r^{E_{0}^{2}})+a_{4}N_{0}\left( 
\frac{\sqrt{c}}{E_{0}^{2}}r^{E_{0}^{2}}\right) .
\end{equation}

\paragraph{Case 3: $E_{0}^{2}>1:$}

For this choice of $E_{0},$equation (37) remains the same and the solution
is given by,%
\begin{equation}
R(r)=a_{5}J_{0}(\frac{\sqrt{c}}{E_{0}^{2}}r^{E_{0}^{2}})+a_{6}N_{0}\left( 
\frac{\sqrt{c}}{E_{0}^{2}}r^{E_{0}^{2}}\right) .
\end{equation}%
Note that $J_{0}\left( x\right) $ and $N_{0}\left( x\right) $ in equations
(40), (42) and (43) are the first kind Bessel and Neumann functions
respectively, with integration constants $a_{i}$ in which $i=1...6.$

Our calculations have revealed that, in general, in each case for
appropriate $a_{i}$, the squared norm $\left\Vert R\right\Vert ^{2}<\infty ,$
which is always square integrable. Hence, the spatial part of the operator
is not essentially self-adjoint for all space $L^{2}(0,\infty )$. Therefore,
the classical singularity at $r=0$ remains quantum singular as well when
probed with massless scalar, bosonic waves.

It is remarkable to note that the considered spacetime in this paper is in
the form of a power - law metrics that can be expressed as%
\begin{equation}
ds^{2}=r^{2E_{0}^{2}}(-dt^{2}+dr^{2})+r^{2}d\theta ^{2}.
\end{equation}%
which satisfies all energy conditions.

Quantum singularity structure of the four - parameter, power - law metrics
in $3+1-$dimensional cylindrically symmetric spacetime has already been
considered by Helliwell and Konkowski (HK) in \cite{Y5}. According to the
classification presented in \cite{Y5}, the metric (35) is Type I with the
only parameters $\beta =2E_{0}^{2}$ and $\gamma =2$ such that $C=1.$ In our
metric the parameter $\beta $ is related to the intensity of the constant
electric field $E_{0}$ so that $\beta =2E_{0}^{2}>0.$ Hence, $\beta <0$ is
not allowed in our study. Furthermore, the absence of an extra dimension in
our study does not provide a wave component with a mode $k$ that has a
crucial effect in the study performed in \cite{Y5}. HK have shown that a
large set of classically singular spacetimes emerges quantum mechanically
non-singular, if it is probed with waves having non - zero azimuthal quantum
number $m$ and axial quantum number $k$. The presence of an extra dimension
in \cite{Y5}, with a parameter $\delta $ is closely related to the geometry
of the spacetime and hence, the considered spacetime in this paper is
different. It is also important to note that the presence of the electric
field in our study increases the rate of divergence in scalar curvature.
This feature amounts to increase the strength of the naked singularity. As a
result, unlike the case considered in \cite{Y5}, the classical naked
singularity remains quantum singular with respect to bosonic wave (spin $0$)
probe.

\subsubsection{Dirac Fields}

In $2+1-$dimensional curved spacetimes, the formalism leading to a solution
of the Dirac equation was given in \cite{39}. This formalism has been used
in \cite{13} and in our earlier studies \cite{14,15,16}. The Dirac equation
in $2+1$ $-$dimensional curved background for a free particle with mass $m$
is given by%
\begin{equation}
i\sigma ^{\mu }\left( x\right) \left[ \partial _{\mu }-\Gamma _{\mu }\left(
x\right) \right] \Psi \left( x\right) =m\Psi \left( x\right) ,
\end{equation}%
where $\Gamma _{\mu }\left( x\right) $\ is the spinorial affine connection
given by

\begin{equation}
\Gamma_{\mu}\left( x\right) =\frac{1}{4}g_{\lambda\alpha}\left[ e_{\nu,\mu
}^{\left( i\right) }(x)e_{\left( i\right) }^{\alpha}(x)-\Gamma_{\nu\mu
}^{\alpha}\left( x\right) \right] s^{\lambda\nu}(x),
\end{equation}

\begin{equation}
s^{\lambda \nu }(x)=\frac{1}{2}\left[ \sigma ^{\lambda }\left( x\right)
,\sigma ^{\nu }\left( x\right) \right] .
\end{equation}%
Since the fermions have only one spin polarization in $2+1-$dimension, the
Dirac matrices $\gamma ^{\left( j\right) }$ can be given in terms of Pauli
spin matrices $\sigma ^{\left( i\right) }$ so that

\begin{equation}
\gamma ^{\left( j\right) }=\left( \sigma ^{\left( 3\right) },i\sigma
^{\left( 1\right) },i\sigma ^{\left( 2\right) }\right) ,
\end{equation}%
where the Latin indices represent internal (local) frame. In this way,

\begin{equation}
\left\{ \gamma ^{\left( i\right) },\gamma ^{\left( j\right) }\right\} =2\eta
^{\left( ij\right) }I_{2\times 2},
\end{equation}%
where $\eta ^{\left( ij\right) }$\ is the Minkowski metric in $2+1-$%
dimension and $I_{2\times 2}$\ is the identity matrix. The coordinate
dependent metric tensor $g_{\mu \nu }\left( x\right) $\ and matrices $\sigma
^{\mu }\left( x\right) $\ are related to the triads $e_{\mu }^{\left(
i\right) }\left( x\right) $\ by

\begin{align}
g_{\mu \nu }\left( x\right) & =e_{\mu }^{\left( i\right) }\left( x\right)
e_{\nu }^{\left( j\right) }\left( x\right) \eta _{\left( ij\right) }, \\
\sigma ^{\mu }\left( x\right) & =e_{\left( i\right) }^{\mu }\gamma ^{\left(
i\right) },  \notag
\end{align}%
where $\mu $\ and $\nu $\ stand for the external (global) indices. The
suitable triads for the metric (21) are given by,

\begin{equation}
e_{\mu }^{\left( i\right) }\left( t,r,\theta \right) =diag\left(
r^{E_{0}^{2}},r^{E_{0}^{2}},r\right) .
\end{equation}%
With reference to our earlier studies in \cite{14,15,16}, the following
ansatz is used for the positive frequency solutions: 
\begin{equation}
\Psi _{n,E}\left( t,x\right) =\left( 
\begin{array}{c}
R_{1n}(r) \\ 
R_{2n}(r)e^{i\theta }%
\end{array}%
\right) e^{in\theta }e^{-iEt}
\end{equation}%
The radial part of the Dirac equations governing the propagation of the
fermionic waves should be examined for a unique self-adjoint extensions for
all space $L^{2}(0,\infty )$. In doing this, the possible values of the
electric field intensity $E_{0}$ should also be taken into consideration. To
consider all these, the behavior of the solution of the radial part of the
Dirac equation $R_{in}\left( r\right) $ $(i=1,2)$ will be investigated near $%
r\rightarrow 0,$ and $r\rightarrow \infty .$

\paragraph{\textbf{The case of }$r$\textbf{\ }$\rightarrow 0:$}

The behavior of the radial part of the Dirac equation for $E_{0}^{2}>1$ is
given by 
\begin{multline}
R_{in}^{\prime \prime }(r)-\frac{\alpha _{1}}{r}R_{in}^{\prime }(r)+\frac{%
\alpha _{1}\left( \alpha _{1}+2\right) }{4r^{2}}R_{in}\left( r\right) =0,%
\text{ } \\
i=1,2
\end{multline}%
in which $\alpha _{1}$ $=E_{0}^{2}-1,$ and the solution is 
\begin{multline}
R_{in}\left( r\right) =b_{1}r^{\frac{\alpha _{1}}{2}+1}+b_{2}r^{\frac{\alpha
_{1}}{2}},\text{ } \\
i=1,2
\end{multline}%
For $E_{0}^{2}=1$ case, the Dirac equation simplifies to 
\begin{multline}
R_{in}^{\prime \prime }(r)+R_{in}^{\prime }(r)-\left[ E^{2}+n(n+1)\right]
R_{in}\left( r\right) =0, \\
\text{\ }i=1,2
\end{multline}%
and the solution is given by%
\begin{multline}
R_{in}\left( r\right) =b_{3}e^{\frac{r\left( -1+\sqrt{1+4\xi }\right) }{2}%
}+b_{4}e^{\frac{-r\left( 1+\sqrt{1+4\xi }\right) }{2}},\text{ \ \ } \\
i=1,2
\end{multline}%
in which $\xi =E^{2}+n(n+1)$ and $E$ is the energy of the Dirac particle.

For $0<E_{0}^{2}<1$ case, the Dirac equation becomes

\begin{multline}
R_{in}^{\prime \prime }(r)+\frac{\beta _{0}}{r}R_{in}^{\prime }(r)+\frac{%
\beta _{0}\left( \beta _{0}+2\right) }{4r^{2}}R_{in}\left( r\right) =0,\text{
} \\
i=1,2
\end{multline}%
in which $\beta _{0}=1-E_{0}^{2},$ and the solution is

\begin{multline}
R_{in}\left( r\right) =b_{5}r^{\frac{E_{0}^{2}+\sqrt{1-4\beta _{0}}}{2}%
}+b_{6}r^{\frac{E_{0}^{2}-\sqrt{1-4\beta _{0}}}{2}},\text{ \ } \\
\text{\ }i=1,2\text{\ \ }
\end{multline}%
such that for a real solution\ the electric field intensity is bounded to $%
\frac{3}{4}\leq E_{0}^{2}<1$. Note that a prime in equations (53), (55) and
(57) denotes a derivative with respect to $r$.

The square integrability of the above solutions corresponding to different
values of electric field intensity $E_{0}$ is checked by calculating the
squared norm given in equation (38). Calculations have indicated that
irrespective of the integration constants $b_{k}$ \ $\left( k=1...6\right) $
and the electric field intensity $E_{0}$, all the solutions obtained near $%
r\rightarrow 0$ are square integrable; that is to say $\left\Vert
R_{in}\right\Vert ^{2}<\infty $.

\paragraph{\textbf{The case of }$r\rightarrow \infty :$}

For the asymptotic case, the radial part of the Dirac equation for the
electric field intensity $E_{0}^{2}>1$ is given by

\begin{multline}
R_{in}^{\prime \prime }(r)+r^{\alpha _{1}}R_{in}^{\prime
}(r)-n(n+1)r^{2\alpha _{1}}R_{in}\left( r\right) =0, \\
\text{\ }i=1,2,
\end{multline}%
whose solution is given by

\begin{multline}
R_{in}\left( r\right) =b_{7}e^{\frac{-r^{E_{0}^{2}\left( \sqrt{1+4\eta }%
+1\right) }}{2E_{0}^{2}}}\times \\
KM(\frac{\left( \alpha _{1}+2\right) \sqrt{1+4\eta }+\alpha _{1}}{2E_{0}^{2}%
\sqrt{1+4\eta }},\frac{\alpha _{1}+2}{E_{0}^{2}},\frac{\sqrt{1+4\eta }%
r^{E_{0}^{2}}}{E_{0}^{2}})r \\
+b_{8}e^{\frac{-r^{E_{0}^{2}\left( \sqrt{1+4\eta }+1\right) }}{2E_{0}^{2}}%
}\times \\
KU(\frac{\left( \alpha _{1}+2\right) \sqrt{1+4\eta }+\alpha _{1}}{2E_{0}^{2}%
\sqrt{1+4\eta }},\frac{\alpha _{1}+2}{E_{0}^{2}},\frac{\sqrt{1+4\eta }%
r^{E_{0}^{2}}}{E_{0}^{2}})r
\end{multline}%
in which $\eta =n(n+1)$, $KM$ and $KU$ stand for $KummerM$ and $KummerU.$

For $E_{0}^{2}=1,$ the behavior of the Dirac equation is

\begin{multline}
R_{in}^{\prime \prime }(r)+R_{in}^{\prime }(r)-mr^{2}R_{in}\left( r\right)
=0, \\
\text{\ }i=1,2,
\end{multline}%
in which $m$ is the mass of the Dirac particle which is taken to be unity
for practical reasons and the solution is given in terms of Kummer function
as,

\begin{multline}
R_{in}\left( r\right) =b_{9}KummerM(\frac{13}{16},\frac{3}{2},r^{2})re^{%
\frac{-r\left( 1+r\right) }{2}}+ \\
b_{10}KummerU(\frac{13}{16},\frac{3}{2},r^{2})re^{\frac{-r\left( 1+r\right) 
}{2}},
\end{multline}%
For $0<E_{0}^{2}<1$ case, the Dirac equation becomes,

\begin{multline}
R_{in}^{\prime \prime }(r)-mr^{2\left( 1-\beta _{0}\right) }R_{in}\left(
r\right) =0,\text{\ } \\
i=1,2,
\end{multline}%
and the solution for $m=1$ and $\beta _{0}=1/2$ is given by%
\begin{equation}
R_{in}\left( r\right) =b_{11}\sqrt{r}I_{1/3}\left( x\right) +b_{12}\sqrt{r}%
K_{1/3}(x)
\end{equation}%
in which $I_{1/3}\left( x\right) $ and $K_{1/3}(x)$ are the first and second
kind modified Bessel functions and $x=-\frac{2r^{3/2}}{3}.$

The obtained solutions in asymptotic case $r\rightarrow \infty ,$ for three
different electric field intensities $E_{0}$ are checked for a square
integrability. Our calculations have revealed that the solutions for $%
b_{8}=b_{10}=0$ and $b_{7}\neq 0,$ $b_{9}\neq 0$ together with $b_{11}\neq 0$
and $b_{12}\neq 0$, the squared norm \ $\left\Vert R_{in}\right\Vert
^{2}\rightarrow \infty $, indicating that the solutions do not belong the
Hilbert space.

From this analysis, we conclude that the radial part of the Dirac operator
is not essentially self-adjoint for all space $L^{2}(0,\infty ),$ and
therefore, the formation of the classical timelike naked singularity in the
presence of the azimuthally symmetric electric field in $2+1-$ dimensional
geometry remains quantum mechanically singular even if it is probed with
spinorial fields.

\subsection{Quantum probes of the radial solution to $\left( 2+1\right) $ $-$%
dimensional gravity coupled to a self-interacting real scalar field}

\subsubsection{Klein - Gordon Fields}

The massless Klein - Gordon equation for the metric (29) after separating
the temporal and spatial parts can be written as%
\begin{equation}
\frac{\partial ^{2}\psi }{\partial t^{2}}=-\mathcal{A}\psi ,
\end{equation}%
in which the spatial operator $\mathcal{A}$ is given by%
\begin{equation}
\mathcal{A=-}K\left( r^{2}\frac{\partial ^{2}}{\partial r^{2}}+2r\frac{%
\partial }{\partial r}+\frac{\partial ^{2}}{\partial \theta ^{2}}\right) .
\end{equation}%
As a requirement of the HM criterion, the spatial operator $\mathcal{A}$
should be investigated for a unique self-adjoint extensions. Hence, it is
required to look for a separable solution to the equation (65), in the form
of $\psi (r,\theta )=R(r)Y(\theta )$ which gives the radial solution as%
\begin{equation}
\frac{\partial ^{2}R\left( r\right) }{\partial r^{2}}+\frac{2}{r}\frac{%
\partial R\left( r\right) }{\partial r}+\frac{1}{r^{2}}\left( c\pm \frac{i}{K%
}\right) R\left( r\right) =0,
\end{equation}%
in which $c$ stands for the separation constant. The essential
self-adjointness of the spatial operator $\mathcal{A}$ requires that neither
of the two solutions of the above equation is square integrable over all
space $L^{2}(0,\infty ).$The square integrability of the solution of the
above equation for each sign $\pm $ is checked by calculating the squared
norm of the obtained solution in which the function space on each $t=$
constant hypersurface $\Sigma $ is defined as $\mathcal{H=}\{R|\left\Vert
R\right\Vert <\infty \}.$ The squared norm for the metric (29) is given by,

\begin{equation}
\left\Vert R\right\Vert ^{2}=\int_{0}^{\infty }\sqrt{\frac{g_{rr}g_{\theta
\theta }}{g_{tt}}}\left\vert R_{\pm }\left( r\right) \right\vert
^{2}dr\simeq \int_{0}^{\infty }\left\vert R_{\pm }\left( r\right)
\right\vert ^{2}dr.
\end{equation}%
We first consider the case when $r\rightarrow 0.$ In this limiting case the
equation (67) simplifies to%
\begin{equation}
\frac{\partial ^{2}R\left( r\right) }{\partial r^{2}}+\frac{1}{r^{2}}\left(
c\pm \frac{i}{K}\right) R\left( r\right) =0,
\end{equation}%
whose solution is%
\begin{equation}
R(r)=C_{1}r^{\delta _{1}}+C_{2}r^{\delta _{2}},
\end{equation}%
in which%
\begin{multline}
\delta _{1}=\frac{1}{2}\left( 1+\sqrt{1-4\upsilon }\right) ,\text{ \ \ \ } \\
\text{\ }\delta _{1}=\frac{1}{2}\left( 1-\sqrt{1-4\upsilon }\right) ,\text{
\ \ \ \ }\upsilon =c\pm \frac{i}{K}.
\end{multline}%
The above solution is checked for square integrability with the norm given
in equation (68). Our analysis have shown that for specific mode of solution
(depending on the value of $\upsilon $) the squared norm diverges i.e. $%
\left\Vert R\right\Vert ^{2}\rightarrow \infty ,$ indicating that the
solution does not belong to Hilbert space.

We consider also the case when $r\rightarrow \infty ,$ so that the equation
(67) approximates to%
\begin{equation}
\frac{\partial ^{2}R\left( r\right) }{\partial r^{2}}+\frac{2}{r}\frac{%
\partial R\left( r\right) }{\partial r}=0,
\end{equation}%
and its solution is given by%
\begin{equation}
R(r)=C_{3}+\frac{C_{4}}{r},
\end{equation}%
in which $C_{k}$ $\left( k=1...4\right) ,$ are the integration constants.
The above solution is square integrable if and only if the integration
constant $C_{4}=0.$ As a consequence, although there exist some modes of
solution that does not belong to the Hilbert space near $r\rightarrow 0$ and 
$r\rightarrow \infty ,$ the generic result is that the considered spacetime
remains quantum singular against bosonic wave probe in view of quantum
mechanics.

\subsubsection{Dirac Fields}

The steps demonstrated previously for the solution of Dirac equation is
applied for the metric given in equation (). The exact radial part of the
Dirac equation is given by%
\begin{multline}
R_{1n}^{\prime \prime }(r)+\frac{1}{r}\left( 1+\frac{E}{M\sqrt{K}r+E}\right)
R_{1n}^{\prime }(r)+ \\
\left[ \frac{1}{r^{2}}\left( \frac{E^{2}}{K}-n^{2}\right) -\frac{En}{%
r^{2}\left( M\sqrt{K}r+E\right) }-M^{2}\right] R_{1n}(r)=0,
\end{multline}%
\begin{multline}
R_{2n}^{\prime \prime }(r)+\frac{1}{r}\left( 1-\frac{E}{M\sqrt{K}r-E}\right)
R_{2n}^{\prime }(r)+ \\
\left[ \frac{1}{r^{2}}\left( \frac{E^{2}}{K}-\left( n+1\right) ^{2}\right) -%
\frac{E\left( n+1\right) }{r^{2}\left( M\sqrt{K}r+E\right) }\right.  \\
\left. -M^{2}\right] R_{2n}(r)=0.
\end{multline}%
The solution to these equations should be investigated for a unique
self-adjoint extensions for all space $L^{2}(0,\infty ).$ This will be done
by considering the behavior of the solution near $r\rightarrow 0$ and $%
r\rightarrow \infty .$

\paragraph{The case when $r\rightarrow 0:$}

The behavior of the radial part of the Dirac equation given in equations
(74) and (75) when $r\rightarrow 0$ are%
\begin{equation}
R_{1n}^{\prime \prime }(r)+\frac{2}{r}R_{1n}^{\prime }(r)-\frac{En}{M\sqrt{K}%
r^{3}}R_{1n}(r)=0,
\end{equation}%
\begin{equation}
R_{2n}^{\prime \prime }(r)+\frac{2}{r}R_{2n}^{\prime }(r)-\frac{E\left(
n+1\right) }{M\sqrt{K}r^{3}}R_{2n}(r)=0,
\end{equation}%
whose solutions are given respectively by%
\begin{equation}
R_{1n}(r)=\frac{d_{1}}{\sqrt{r}}J_{1}\left( x_{1}\right) +\frac{d_{2}}{\sqrt{%
r}}N_{1}\left( x_{1}\right)
\end{equation}%
\begin{equation}
R_{2n}(r)=\frac{d_{3}}{\sqrt{r}}J_{1}\left( x_{2}\right) +\frac{d_{4}}{\sqrt{%
r}}N_{1}\left( x_{2}\right)
\end{equation}%
where $d_{k}$ $\left( k=1...4\right) $ are the integration constants, $%
J_{1}\left( x_{i}\right) $ and $N_{1}(x_{i})$ are the Bessel and Neumann
functions with order the $1$ such that $x_{1}=2\sqrt{\frac{-\lambda _{1}}{r}}
$ and $x_{2}=2\sqrt{\frac{-\lambda _{2}}{r}}$, $\lambda _{1}=\frac{En}{M%
\sqrt{K}}$ , $\lambda _{2}=\frac{E\left( n+1\right) }{M\sqrt{K}}$. The
square integrability of these solutions is checked with the norm given in
equation (). The outcome of our analysis is that none of the obtained
solutions are not belong to the Hilbert space. In other words the squared
norm $\left\Vert R\right\Vert ^{2}\rightarrow \infty .$

\paragraph{The case when $r\rightarrow \infty :$}

The behavior of the radial part of the Dirac equation given in equations
(74) and (75) when $r\rightarrow \infty $ are%
\begin{equation}
R_{1n}^{\prime \prime }(r)+\frac{1}{r}R_{1n}^{\prime }(r)-\frac{En}{M\sqrt{K}%
r^{3}}R_{1n}(r)=0,
\end{equation}%
\begin{equation}
R_{2n}^{\prime \prime }(r)+\frac{2}{r}R_{2n}^{\prime }(r)-\frac{E\left(
n+1\right) }{M\sqrt{K}r^{3}}R_{2n}(r)=0,
\end{equation}%
whose solutions are given respectively by%
\begin{equation}
R_{in}(r)=l_{i}J_{0}\left( x_{3}\right) +l_{i}N_{0}\left( x_{3}\right) ,%
\text{ \ \ \ \ }i=1,2
\end{equation}%
in which $l_{i}$ $\left( i=1,2\right) $ are the integration constants, $%
J_{0}\left( x_{3}\right) $ and $N_{0}\left( x_{3}\right) $ are the Bessel
and Neumann functions with order $0$ such that $x_{3}=\sqrt{-M}r$. The
analysis for square integrability have shown that this solution is not
square integrable. Alternatively, $\left\Vert R\right\Vert ^{2}\rightarrow
\infty .$

As a result of this analysis, the radial part of the Dirac operator on this
spacetime is essentially self adjoint and therefore, the timelike naked
singularity in the $\left( 2+1\right) -$dimensinal gravity sourced by a real
scalar field is quantum mechanically wave regular. This result indicates
that the spin of the wave is effective in healing the singularity.

\section{CONCLUSION}

In this paper, the formation of timelike naked singularities in $2+1-$%
dimensional power - law spacetimes in linear Einstein-Maxwell and Einstein -
scalar theories powered by azimuthally symmetric electric field and a
self-interacting real scalar field, respectively, are investigated from
quantum mechanical point of view. \ Two types of waves with different spin
structures are used to probe the timelike naked singularities that develops
at $r=0$.

We showed that the scalar ( bosonic, spin $0$) wave probe is not effective
in healing the classical timelike naked singularities formed both in $2+1-$%
dimensional power - law metrics sourced by azimuthally symmetric electric
field and a real scalar field. From the Kretschmann scalar given in (22), it
is easy to observe that, the azimuthally symmetric electric field $E_{0}$,
serves to increase the rate of divergence and hence, in some sense yields a
stronger naked singularity at $r=0$.

It is important to compare our study with that of HK in \cite{Y5} and of BFW
in \cite{Y1}, where the occurrence of timelike naked singularities are
analyzed in quantum mechanical point of view. First, we compare with the
work of HK. The metric considered by HK is a four-parameter power - law
metrics in $3+1$ $-$dimensions, whose metric coefficients behave as power
laws in the radial coordinate $r$, in the small $r$ approximation. The
common point of our study with that of HK is that both metrics are in the
power - law form. However, there is a significant differences between our
metrics and that of HK. Although, the duality of the Maxwell field $2-$form
in $3+1-$dimensions is still a $2-$form, but in $2+1-$dimension, duality of
the Maxwell field maps a $2-$form into $1-$form or vice versa. Another
distinction is that the presence of an extra dimension in $3+1-$dimension,
allows the mode solution in the wave equation to depend on the axial quantum
number $k,$ that has a crucial effect on healing the singularity. We may add
also that $2+1-$dimensional spacetime is a brane in $3+1-$dimensions and
therefore dilutation of the gravitational singularity in higher dimensions
is not unexpected. In view of all these, the spacetimes considered in this
paper and that of HK is topologically different. Another interesting result
of this study is that, all the energy conditions WEC, DEC, SEC and the
causality conditions are satisfied, but the spacetime is quantum singular.
There was an attempt to relate the quantum regular/singular spacetimes with
the energy conditions in \cite{Y5}. The result obtained in our case
indicates that eliminating the quantum singular spacetimes by just invoking
the energy conditions is not a reliable method. This result confirms the
comment made by HK in \cite{Y5}, that invoking energy conditions is not
guaranteed physically.

On the other hand, the metric considered by BFW in \cite{Y1} is a
spherically symmetric power - law metrics with dimensions $n\geq 4.$ Here,
the timelike naked singularity is probed with scalar field and shown that
the resulting spacetime is quantum singular. The possible connection with
the energy conditions on the quantum resolution of the timelike naked
singularity is also addressed in \cite{Y1}. The general conclusion drawn for 
$n\geq 4$ dimensional spherically symmetric power - law metrics is that " 
\textit{metrics with timelike singularities of power -law type satisfying
the strict Dominant Energy Condition remain singular when probed with scalar
waves} ". \ Although this statement is in conform with our results for the
case of scalar wave probe, what we believe is that it can not be taken as a
general rule.

A contradicting result to above statement is obtained for the exact radial
solution with a self-interacting, real, scalar field coupled to the $\left(
2+1\right) $ $-$dimensional gravity. The timelike naked singularity is
probed with Dirac field (fermionic, spin $1/2$) that obeys the Dirac
equation. We showed that the spatial radial part of the Dirac operator has a
unique extension so that it is essentially self-adjoint. And as a result,
the classical timelike naked singularity in the considered spacetime remains
quantum regular when probed with fermions.

The notable result obtained in this paper, and also in earlier studies along
this direction has indicated that, the quantum healing of the classically
singular spacetime crucially depends on the wave that we probe the
singularity. In order to understand the generic behavior of the $2+1-$%
dimensional spacetimes, more spacetimes should be investigated. So far,
vacuum Einstein, Einstein-Maxwell (both linear and nonlinear) and
Einstein-Maxwell-dilaton solutions are investigated. As a future research,
timelike naked singularities in $2+1-$dimensional Einstein-scalar (minimally
coupled) and Einstein-Maxwell-scalar solutions should also be investigated
from quantum mechanical point of view. It will be interesting also to
consider the spinorial wave generalization of the power - law metrics
considered in \cite{Y5,Y1}.

\bigskip

\end{document}